\documentclass[12pt]{article}
\newcommand{\be}{\begin{equation}}
\newcommand{\ee}{\end{equation}}
\newcommand{\bea}{\begin{eqnarray}}
\newcommand{\eea}{\end{eqnarray}}
\newcommand{\nn}{\nonumber \\}
\newcommand{\p}[1]{(\ref{#1})}
\newcommand\tb{\bar\theta}
\newcommand{\Qb}{\bar{Q}}
\newcommand{\Sb}{\bar S}
\newcommand{\Db}{\bar{D}}
\newcommand{\cD}{{\cal D}}
\def\theequation{\arabic{section}.\arabic{equation}}
\begin{document}
\begin{titlepage}
\begin{flushright}
JINR E2-99-305\\
LPENSL-TH-21/99\\
hep-th/9912222 \\
December 1999
\end{flushright}
\vskip1cm
\centerline{\large\bf 1/4 Partial Breaking of Global Supersymmetry}
 \centerline{\large\bf and New Superparticle Actions}

\vskip1cm
\centerline{\bf F. Delduc${}^{\, a,1}$, E. Ivanov${}^{\, b,2}$, S.
Krivonos${}^{\, b,3}$}

\bigskip
\centerline{$^a${\it Laboratoire de Physique}$^{\,
\dagger }$, {\it Groupe de Physique Th\'eorique ENS Lyon}}
\centerline{\it 46, all\'ee d'Italie, F - 69364 - Lyon CEDEX 07}
\bigskip
\centerline{$^b${\it Bogoliubov Laboratory of Theoretical Physics, JINR,}}
\centerline{\it 141 980, Dubna, Moscow Region, Russian Federation}

\vskip1cm

\begin{abstract}
\noindent We construct  the worldline superfield
massive superparticle actions which preserve $1/4$ portion of
the underlying higher-dimensional supersymmetry. The rest of supersymmetry
is spontaneously broken and realized by nonlinear transformations.
We consider  the cases of $N=4 \rightarrow N=1$ and  $N=8 \rightarrow N=2$
partial breaking. In the first case we  present the corresponding
Green-Schwarz type target superspace action with one $\kappa$-supersymmetry.
It is related to the superfield action via a field redefinition. In the
second case we find out two possible models, one of which is a direct
generalization of the $N=4 \rightarrow N=1$  case, while another is
essentially different. For the first model we formulate Green-Schwarz type 
action
with two $\kappa$-supersymmetries. We elaborate on the bosonic part of
the superfield action for the second model and find  that only in
two special limits it takes the standard Nambu-Goto form.
In the general case it is determined by a fourth-order algebraic equation.
The characteristic common feature of  these new superparticle models is
that  the algebras of their spontaneously broken supersymmetries are
non-trivial truncations of the general extensions of
$N=1$ and  $N=2$ Poincar\'e  $D=4$ superalgebras  by tensorial
central charges.
\end{abstract}
\vfill
{\it E-Mail:}\\
{\it 1) delduc@ens-lyon.fr}\\
{\it 2) eivanov@thsun1.jinr.ru}\\
{\it 3) krivonos@thsun1.jinr.ru}
\nopagebreak
\begin{flushleft}
\rule{2in}{0.03cm} \\
{\footnotesize \ ${}^{\dagger}$
UMR 8514, CNRS et \'Ecole Normale Sup\'erieure de Lyon.}
\end{flushleft}

\newpage

\end{titlepage}
\section{Introduction}
The most attractive feature of the
description of superbranes based upon the concept of
partial spontaneous breaking of global supersymmetry (PBGS)
\cite{hp} - \cite{ik} is the manifest off-shell realization of the
corresponding worldvolume supersymmetry. In this approach, the
physical worldvolume multiplets of the given superbrane are
represented by superfields given on the proper worldvolume superspace.
They are interpreted as Goldstone superfields realizing spontaneous
breaking of the full brane supersymmetry group down to its unbroken
worldvolume subgroup. The spontaneously broken supersymmetry is
realized on the Goldstone superfields by inhomogeneous and nonlinear
transformations. The invariant Goldstone superfields actions, after
passing to components and, in general, eliminating auxiliary
fields by their equations of motion, coincide with gauge-fixed forms
of the relevant Green-Schwarz type actions.

Until now, only the examples
of $1/2$  breaking of supersymmetry corresponding to the
standard  BPS  $p$-branes and D-branes were treated in the
literature on the PBGS. On the other hand, recently there was some
interest  in the $1/4$ and other unusual fractional PBGS options,
mainly caused by the
existence of  the $D=11$ supergravity solutions which preserve various
fractions of the underlying 32 supersymmetries and admit a nice
interpretation in terms of intersecting  branes  (see, e.g.,
\cite{bdl}-\cite{gh} and references therein).

In view of this, it seems interesting to
extend the manifestly worldvolume supersymmetric PBGS description
of branes to the $1/4$ breaking  and other fractional
patterns. In the present  paper, as first steps toward  this goal, we
consider several  examples of the $1/4$  PBGS superfield actions in the
simplest case of massive superparticle.

The construction of the Goldstone superfield action is the most
difficult part of the PBGS approach. The generic methods of nonlinear
realizations \cite{1}-\cite{3} which nicely work in the case
of standard internal symmetry and
space-time groups prove to be not too helpful when trying to employ
them for constructing PBGS invariants. All the known
Goldstone superfield Lagrangians are of the Chern-Simons or WZNW type,
in  the sense that they are not tensors with respect to the hidden
supersymmetry transformations. The latter shift them by a full
derivative, thus leaving the action invariant up to surface terms.
As the result, one cannot directly apply the powerful method of
covariant Cartan forms for constructing invariant actions.

A way around this difficulty was proposed in \cite{{bg2},{bg3}} and
used to construct the $N=1$, $D=4$ superfield
actions providing the PBGS description of $N=2$ D3-superbrane and super
3-brane.
It is based on the idea of embedding the basic Goldstone superfield into
some {\it linear} multiplet of the underlying supersymmetry  group.
Initially this multiplet comprises
a set of independent worldvolume superfields. After imposing appropriate
covariant constraints one succeeds in expressing all these superfields
in terms of the basic Goldstone ones. One of the superfields of the
initial linear representation is shifted by a full derivative
under the broken supersymmetry transformations and so can be chosen as the
Goldstone superfields Lagrangian.

This procedure was further elaborated and applied to other cases of
PBGS in refs. \cite{rt}, \cite{gpr}, \cite{ik}. However, its generality
and usefulness are obscured by the fact that choosing the
linear representation to start with and picking up appropriate
constraints are a sort of guess-work. This guess-work becomes
rather intricate in more complicated cases like the $1/4$ PBGS
patterns we are interested in here.

In order to construct the Goldstone superfield actions for this case we
propose a more systematic version of the above procedure,
following the general  method of setting up linear representations of
supersymmetry in  terms of the appropriate nonlinear realizations
\cite{{ikap1},{ikap2}}.  Though this method was worked out originally for the
case of total  breaking of supersymmetry, it can be straightforwardly extended
to  the PBGS case. Its main advantage consists in that it allows one to
avoid seeking for the constraints on the original linearly transforming
superfields. They are expressed through the Goldstone superfields by
making use of some algorithmic algebraic computation.

In Sect. 2 we explain this method on the toy examples of the $N=2
\rightarrow N=1$ and $N=4 \rightarrow N=2$ PBGS in $d=1$ which
were studied earlier from a different point of view in
\cite{{town},{town2},{gauntY},{iak}}. Then in
Sect. 3 we apply it to construct  a new model of massive  superparticle with
the  $1/4$ PBGS pattern $N=4 \rightarrow N=1$. We find that the  corresponding
$N=4$, $d=1$ superalgebra does not
follow  by a dimensional reduction from the standard $N=1$ Poincar\'e
superalgebra in $D=4$. It is an extension of the latter by the Lorentz
tensorial central charges (see \cite{{1},{ferrp},{gh},{blstenz}} and references 
therein)
which yields this $N=4$, $d=1$  superalgebra via dimensional reduction. In
Sect. 4 we give the  corresponding Green-Schwarz type action revealing just
one local $\kappa$-symmetry and establish the precise correspondence between
the  PBGS and Green-Schwarz actions. In Sect. 5 we construct, using
analogous methods, the superparticle model realizing the PBGS
$N=8 \rightarrow N=2$. We find that there exist two versions of
such model. They differ in the superfield contents and in the
structure of the physical bosons action. Only in the first case
this action has the standard static gauge Nambu-Goto form of the massive
particle action. We present the Green-Schwarz type action for this
case, with two independent $\kappa$ symmetries.

\setcounter{equation}0
\section{Two examples of 1/2 PBGS in d=1}
\subsection{N=2 $\rightarrow$
N=1} To explain the key
features of our approach we start from a simplest example of
$N=2 \rightarrow N=1$ partial breaking of
global supersymmetry in $d=1$.

The anticommutation relation of $N=1, d=1$ Poincar\'e superalgebra
reads
\be
\left\{ Q,Q \right\}= 2P~. \label{n1susy}
\ee
We wish to construct a $N=1$ superfield action
which is also invariant with respect to one extra
spontaneously broken $d=1$ supersymmetry. Thus our basic
objects will be superfields given on
$N=1, d=1$ superspace with the coordinates
 $\left\{ t,\theta \right\}, ( {\bar t}=-t,\bar\theta=\theta )$
\footnote{Henceforth, we use the following convention for
the evolution parameter and action: $t = it'$ and
$S = iS'$, where $t'$ and $S'$ are the standard real time and action.}

Let us define the fermionic and bosonic
superfields $\psi(t,\theta )$ and $v (t,\theta )$ related as
\be \label{n1psirho}
\psi =\frac{1}{2} \cD v \;, \qquad ( \bar\psi=\psi,\; \bar v=-v )~,
\ee
where $\cD $ is the spinor covariant derivative
\be
{\cal D}=\frac{\partial}{\partial \theta}+\theta\partial_{t}\; ,\qquad
\left\{ \cD, \cD \right\} =2\partial_{t}~.
\ee
If we now introduce an additional, arbitrary for the moment, real
spinor superfield $\Upsilon (t,\theta )$ then it becomes possible to
realize an extra $N=1, d=1$ supersymmetry on the spinors $\psi $ and
$\Upsilon$. Assuming the second supersymmetry to be spontaneously
broken, the linear  transformation laws can be written as
\be\label{n1tr1}
\delta\psi=\epsilon\left( 1- \cD \Upsilon\right)\;, \qquad
\delta\Upsilon=\epsilon \cD \psi~.
\ee
The presence of the constant shift in the transformations \p{n1tr1} just
means the spontaneous breaking of this supersymmetry and suggests the
interpretation of $\psi$ as the Goldstone
fermionic superfield  accompanying the linear realization of
this breaking.
One can check that the generator $S$ of the
transformations \p{n1tr1} forms the $N=1,d=1$ superalgebra
\be
\left\{ S,S \right\}= 2P  \label{n1susy1}
\ee
and anticommutes with the generator $Q$ of the manifest $N=1$ supersymmetry
\p{n1susy} in the given realization.\footnote{We should stress that
hereafter the superalgebras we deal with are those of the
superfield variations. It is sufficient to consider them for our aim of
constructing invariant actions. The superalgebras of charges
and supercharges constructed from the PBGS actions by N\"other procedure
are different: they inevitably include some constant central
charges which are crucial for evading \cite{{hp},{hlp}} the famous Witten's
no-go  theorem \cite{Witten}. In particular, such a central charge
will appear as a shift of
$P$  in the r.h.s. of the anticommutator \p{n1susy1} of generators of the
spontaneously broken $S$ supersymmetry. These central charges do not
produce any transformation on the fields, as opposed to another,
``active'' sort  of central charges which generate actual symmetries
and  are of heavy use throughout this paper.}
However, from \p{n1tr1} one can also extract the transformation law of the
scalar superfield $v$:
\be \label{n1tr1a}
\delta v =  -2\epsilon\left( \theta -\Upsilon\right)\; ,\qquad
\delta\Upsilon = \frac{1}{2}\epsilon{\partial_t  v}~.
\ee
Now, the brackets of the manifest and second $N=1$
supersymmetries yield a constant shift of the superfield $v$,
which means that a central charge $Z$ appears in the anticommutators
of these two $N=1$ supersymmetries and $v$ is the bosonic Goldstone
superfield associated with the broken central charge
transformations
\be
\left\{ Q, S \right\} = 2Z~. \label{n1susy2}
\ee

Thus, as long as we limit our attention to the Goldstone fermionic superfield
only, the full supersymmetry algebra is given by \p{n1susy}, \p{n1susy1} with
no need  for central charge. Imposing the relation \p{n1psirho}
introduces the active central charge as in \p{n1susy2} with $v$ as the
corresponding Goldstone superfield. As we shall see soon, these options
lead to PBGS theories with different physical fields contents.

Let us note that the field $\Upsilon$
is a good candidate for the Lagrangian density. Indeed, the integral
\be\label{n1action}
I=\int dt d\theta \;\Upsilon
\ee
is manifestly invariant with respect to the first $N=1$ supersymmetry.
It is also invariant with respect to the second supersymmetry because
the integrand is shifted by a spinor derivative under the variation
\p{n1tr1}.

If the superfield $\Upsilon$ is regarded as independent, the ``action''
\p{n1action} is of course meaningless. It becomes meaningful after
expressing $\Upsilon$ in terms of the Goldstone superfield $\psi $.
This can be performed in a covariant way, resorting to the method
of \cite{{ikap1},{ikap2}}.

As the first step one finds the finite transformations of the second
supersymmetry for the basic superfields $\psi, \Upsilon$ which for the case
at hand just coincide with \p{n1tr1}.

Next, one introduces the Goldstone fermion of nonlinear realization
$\eta$ having the following transformation law under the
second supersymmetry:
\be\label{n1tr1b}
\delta\eta=\epsilon +\epsilon\eta \partial_t \eta~.
\ee
This object, together with $N=1$ superspace coordinates, parametrize
the supersymmetry group manifold. The infinitesimal transformation law
\p{n1tr1b} correspond to the standard group composition law
for the exponential parametrization of the supergroup element.

The crucial step is to substitute $( -\eta )$ for
the parameter $\epsilon$ in the r.h.s. of \p{n1tr1}
and to define new superfields
\be\label{n1tr3}
\tilde\psi=\psi -\eta\left( 1- \cD \Upsilon\right)\;, \qquad
\tilde\Upsilon=\Upsilon-\eta \cD \psi~.
\ee
One can check that the newly defined superfields
$\tilde{\psi},\tilde{\Upsilon}$
transform homogeneously and {\it independently} of each other under the
second spontaneously broken supersymmetry \cite{ikap1}:
\be
\delta \tilde\psi = \epsilon\eta\partial_t \tilde\psi\; , \qquad
\delta \tilde\Upsilon = \epsilon \eta \partial_t \tilde\Upsilon \; .
\ee
Thus it is the covariant operation to put these superfields equal
to zero\footnote{In principle, one could choose an arbitrary constant
as the r.h.s. of the second constraint. This does not influence our further
consideration.}
\be\label{n1neweq}
\tilde{\psi} =0~, \qquad \tilde{\Upsilon}=0~.
\ee
These constraints imply the following set of  equations
for the original superfields $\psi,\eta,\Upsilon$
\be\label{n1eq}
\psi-\eta\left( 1-\cD \Upsilon \right)=0\;, \qquad
\Upsilon-\eta \cD \psi =0~,
\ee
providing the manifestly covariant way of expressing them in terms
of a few basic ones. As such one can take either the Goldstone fermion of
the nonlinear realization, or that of the linear realization.
The solution to eqs. \p{n1eq} is given by
\bea
&& \Upsilon=\frac{2\psi \cD \psi}{1+\sqrt{1-4\cD\psi\; \cD\psi}} =
\frac{\eta\cD\eta}{1+\cD \eta \cD \eta} \;, \label{n1phi} \\
&& \eta=\frac{2\psi}{1+\sqrt{1-4\cD \psi \cD\psi}}~. \label{n1gf}
\eea
We see that eq. \p{n1phi} gives the expression of $\Upsilon$  through the
Goldstone superfields, while eq. \p{n1gf}
is just the equivalency relation between  the two types of the Goldstone
fermionic superfields.

We should stress that the transformation properties of the
Goldstone fermion  $\eta $ \p{n1tr1b} actually follow
from their expressions \p{n1gf} and the known
transformation properties of $\psi,\Upsilon$ \p{n1tr1}. Therefore
we do not need to know them beforehand.

Finally, after the substitution of the expression \p{n1phi}
and using \p{n1psirho} the integral \p{n1action} yields
the sought Goldstone superfield action
\be\label{n1action1}
S_v= \int dt d \theta\, \Upsilon \equiv \frac{1}{2}
\int dt d \theta \,
\frac{ \partial_t v \cD v}{1+\sqrt{1-\left(\partial_t v \right)^2}}~.
\ee
For the physical bosonic component $v|_{\theta=0}$
one obtains just the $D=2$ massive particle action in the static
gauge:
\be
S_v^{bos}=\frac{1}{2}\int dt
\left(1-\sqrt{1-\left( \partial_t v\right)^2}\right)~.
\ee
Thus, the action \p{n1action1} can be naturally interpreted as a
manifestly $N=1,d=1$ supersymmetric worldline action of the superparticle
in $D=2$.

Note that the action \p{n1action1} is related by a simple field
redefinition to the $D=2$ superparticle action which was constructed in
\cite{town2} proceeding from the standard nonlinear realizations
approach. It was also obtained in \cite{iak} as the low-energy
collective coordinates action in some two-dimensional solitonic models.
We used this system just as a simple illustration of our approach to the
construction of the PBGS Goldstone superfields actions.

Finally, we would like to mention that there exists
one more possibility for the Goldstone superfield action. Namely, one can
consider $\psi$ as an independent superfield, without
assuming the ``prepotential'' representation \p{n1psirho}.
This choice corresponds to the $N=2$ supersymmetry
algebra  with $Z=0$ in \p{n1susy2}. In this case the
off-shell wordline Goldstone  multiplet consists of the physical fermion
$\psi\vert_{\theta = 0}$ and an auxiliary bosonic field
$\cD \psi\vert_{\theta = 0}$ , i.e. it includes no physical boson at all.
Such an exotic  structure
is of course an artifact of the one-dimensional case where  supersymmetry
in general does not require matching of the on-shell fermionic and bosonic
degrees of freedom. The corresponding Goldstone  action
is still  given by \p{n1action1}, with $\psi$ standing
for $\frac{1}{2}\cD v$. The auxiliary field $\cD \psi$
vanishes by its equations of motion like in the free case (with
$\psi \cD \psi $  as the Lagrangian), and the action of
the physical fermion turns out to be just the $N=1, d=1$ Volkov-Akulov
action \cite{va}. It corresponds  to the total breaking of $N=1, d=1$ $S$
supersymmetry. Any  trace of unbroken $Q$ supersymmetry \p{susy}
disappears on shell.  Moreover, this $d=1$ Volkov-Akulov action coincides
with the free one.

\subsection{N=4 $\rightarrow$ N=2}

For completeness we briefly consider one more
simple example of PBGS in $d=1$, namely the
$N=4 \rightarrow N=2$ one.

The $N=2, d=1$ Poncar\'e superalgebra without central charges reads
\be
\left\{ Q,Q \right\}=
\left\{ \Qb, \Qb \right\} = 0\; , \qquad
\left\{ Q,\Qb \right\} = P~. \label{susy}
\ee
Our basic objects will be the fermionic and bosonic
$N=2, \; d=1$ superfields $\psi(t,\theta ,\tb),
\bar\psi (t,\theta ,\tb)$
and $\rho (t,\theta ,\tb)$ where $\left\{ t,\theta,\tb \right\}$
are $N=2, d=1$ superspace coordinates.
We assume that $\rho$ is a real superfield
while $\psi, \bar\psi$ are related to it by
\be \label{psirho}
\psi =-\frac{1}{2} \Db\rho~, \qquad \bar\psi = \frac{1}{2} D \rho~,
\ee
where the spinor covariant derivatives are defined by
\be
D=\frac{\partial}{\partial \theta}+\frac{1}{2}\tb\partial_{t}\; , \quad
\Db=\frac{\partial}{\partial \tb}+\frac{1}{2}\theta\partial_{t}\; , \quad
\left\{ D, \Db \right\} =\partial_{t} \; , \quad
 D^2 =\Db{}^2 = 0~.
\ee
By construction, the fermionic superfields are chiral (antichiral)
\be \label{chirps}
\Db\psi = 0~, \qquad   D \bar\psi = 0~.
\ee

After introducing an additional  scalar
superfield $\Phi(t,\theta,\tb)$  one can realize an extra
$N=2, d=1$ supersymmetry on the spinors $\psi,\bar\psi$ and scalar $\Phi$
(the second supersymmetry with the generators $S,\Sb$ is supposed to be
spontaneously broken):
\be\label{tr1}
\delta\psi=\epsilon\left( 1 -\Db D\Phi\right)\;, \quad
\delta{\bar\psi}={\bar\epsilon}\left( 1+D\Db \Phi\right)\;, \quad
\delta\Phi= \epsilon {\bar\psi} -{\bar\epsilon} \psi~.
\ee
The transformation law of the
scalar superfield $\rho$ reads:
\be \label{tr1a}
\delta \rho = 2\left(\theta {\bar\epsilon} -\tb \epsilon +  D\Phi\epsilon
 + \Db \Phi{\bar\epsilon}\right)~.
\ee
Once again, due to the explicit presence of $\theta,\tb$ in \p{tr1a},
the brackets of the manifest and second $N=2$ supersymmetries
yield a constant shift of the superfield $\rho$, and therefore a central
charge $Z$ appears in the anticommutators of these two
$N=2$ supersymmetries
\footnote{This multiplet is a  $d=1$ reduction of chiral $N=1\;,\; D=4$
multiplet.}
\be
\left\{ S, \Sb \right\} = P\; , \qquad
\left\{ Q, \Sb \right\} = 2Z\; , \qquad
\left\{ \Qb, S \right\} = 2Z \;. \label{susy2}
\ee

The field $\Phi$ can be treated as the Lagrangian density and the
action
\be\label{action}
S=\int dt d^2\theta \;\Phi
\ee
is manifestly invariant with respect to both $N=2$ supersymmetries.
To express $\Phi$ in terms of the Goldstone superfields $\psi,\bar\psi$
we use the same method \cite{{ikap1},{ikap2}} as in the previous subsection.

The finite transformations of the second
supersymmetry for the basic superfields $\psi,\bar\psi, \Phi$
can be easily computed
\bea\label{tr2} &&\Delta\psi=\epsilon\left(
1 -\Db D\Phi\right)+\frac{1}{2}\epsilon{\bar\epsilon}\partial_t\psi\;, \qquad
\Delta{\bar\psi}={\bar\epsilon}\left( 1+D\Db \Phi\right)-
      \frac{1}{2}\epsilon{\bar\epsilon}\partial_t{\bar\psi}\;, \nn
&&\Delta\Phi=\epsilon{\bar\psi}-{\bar\epsilon} \psi+
          \epsilon{\bar\epsilon}\left( 1+\frac{1}{2}\left[
D,\Db\right]\Phi \right)~.
\eea

Then one introduces the Goldstone fermions of nonlinear realization
$\eta,\bar\eta$ with the following transformation laws under the
second supersymmetry:
\be\label{tr1b}
\delta\eta=\epsilon+\frac{1}{2}
\left(\epsilon{\bar\eta}+{\bar\epsilon}\eta\right)\partial_t \eta~,
\qquad
\delta{\bar\eta}=\bar\epsilon+\frac{1}{2}\left(\epsilon{\bar\eta}+{\bar\epsilon}
\eta\right)
            \partial_t \bar\eta~,
\ee
and substitutes $(-\eta,-\bar\eta )$ for
the parameters $(\epsilon,\bar\epsilon )$ in the r.h.s. of \p{tr2}
to define the new superfields
\bea\label{tr3}
&&\tilde{\psi}=\psi-\eta\left(
1- \Db D\Phi\right)+\frac{1}{2}\eta{\bar\eta}\partial_t\psi\;, \qquad
\tilde{\bar\psi}={\bar\psi}-{\bar\eta}\left( 1+D\Db \Phi\right)-
\frac{1}{2}\eta{\bar\eta}\partial_t{\bar\psi}\;, \nn
&&\tilde{\Phi}=\Phi-\eta{\bar\psi}+{\bar\eta} \psi +
   \eta{\bar\eta}\left( 1+\frac{1}{2}\left[ D,\Db\right]\Phi
\right)~.  \eea
Like in the previous example they transform independently of each another
and homogeneously under the spontaneously broken supersymmetry.

As the last step one puts these superfields equal to zero
\be\label{neweq}
\tilde{\psi} =0~,\qquad \tilde{\bar\psi}=0~, \quad \tilde{\Phi}=0~.
\ee
The solution to eqs. \p{neweq} is given by
\bea
&& \Phi=\frac{2\psi\bar\psi}{1+\sqrt{1-4 D \psi\; \Db{\bar\psi}}} =
\frac{\eta\bar\eta}{1+\Db D\left(\eta\bar\eta\right)} \; , \label{phi} \\
&& \eta=\frac{\psi}{1-\Db D\Phi}+\frac{\frac{1}{2}\psi{\bar\psi}\partial_t \psi}
       {\left( 1-\Db D\Phi\right)^3} \; , \quad
 \bar\eta=\frac{\bar\psi}{1+D\Db \Phi}-\frac{\frac{1}{2}\psi{\bar\psi}
\partial_t \bar\psi}
       {\left(1+ D\Db \Phi\right)^3}~. \label{gf}
\eea
Note that  the Goldstone fermions
$\eta,\bar\eta$ obey the covariant chirality conditions:
\be
\Db\eta+\frac{1}{2}\eta \Db{\bar\eta} \partial_t \eta =0 \;, \qquad
D\bar\eta+\frac{1}{2}\bar\eta D \eta \partial_t \bar\eta =0
\ee
which are equivalent to the ordinary chirality conditions \p{chirps}
for $\psi,\bar\psi$.

After the substitution of the expression \p{phi}
and using \p{psirho} the action \p{action} takes the form
\be\label{action1}
S_\rho= \int dt d^2 \theta\, \Phi \equiv -\frac{1}{2}
\int dt d^2 \theta \,
\frac{ \Db\rho D\rho}{1+\sqrt{1+\left(D \Db \rho\right)
\left(\Db D\rho\right)}}\;.
\ee
For the physical bosonic component $\rho|_{\theta=0}$
one obtains just the $D=2$ massive particle action in the static
gauge\footnote{ We use the convention $\int d^2\theta \equiv \int \Db D$.}:
\be
S_\rho^{bos}=\frac{1}{2}\int dt \left( 1- \sqrt{1+(\partial_t \rho)^2} \right)
~.
\ee
Thus, the action \p{action1} can be naturally interpreted as a
manifestly $N=2,d=1$ supersymmetric worldline action of the
superparticle in $D=2$. Note that the auxiliary field in $\rho$ is
vanishing on shell.

Like in other PBGS theories, in our case
the Goldstone fermions can be placed into different multiplets
of unbroken $N=2,\; d=1$ supersymmetry. Instead of the real scalar
superfield we can choose  chiral-anti-chiral bosonic superfields
$\phi,\bar\phi$ as the basic Goldstone ones:
\be\label{chir1}
\psi=-\frac{1}{2}\Db \phi ,
\qquad \bar\psi=\frac{1}{2} D {\bar\phi}, \qquad D\phi=\Db{\bar\phi}=0\;.
\ee
The transformation properties of $\phi,\bar\phi$ with respect to
the spontaneously broken supersymmetry are given by
\be\label{tr33}
\delta\phi=2 \epsilon\left( {\bar\theta}-D\Phi\right)\;, \quad
\delta{\bar\phi}=-2{\bar\epsilon}\left( \theta+\Db \Phi\right)\;, \quad
\delta\Phi=\frac{1}{2}\left( \epsilon D\bar\phi+
     {\bar\epsilon} \Db \phi\right)~.
\ee
 From \p{tr33} one finds that  the brackets of manifest and
spontaneously broken supersymmetry contain a complex central charge
$Z$:
\be
\left\{ Q, \Sb \right\} = 2Z\; , \qquad
\left\{ \Qb, S \right\} = 2 {\bar Z} \;, \label{susy33}
\ee
which is realized as shifts of $\phi,\bar\phi$.

Substituting the expressions for $\psi,\bar\psi$ \p{chir1} into
\p{phi}, one finds the action
\be\label{actionm}
S_{\phi} = -\frac{1}{2}\int dt d^2\theta
\,\frac{ \Db \phi  D{\bar\phi}}{1+\sqrt{1+\partial_t
      \phi\partial_t {\bar\phi} }}~.
\ee
Its bosonic core is the
$D=3$ particle action in the static gauge:
\be\label{vm}
S_\phi^{bos}=\frac{1}{2}\int dt \left( 1- \sqrt{1+\partial_t \varphi
     \partial_t{\bar\varphi}  }  \right) \;,
\ee
where $\varphi=\phi|_{\theta=0},{\bar\varphi}={\bar\phi}|_{\theta=0}$.
Therefore, \p{actionm} is a
manifestly $N=2,d=1$ supersymmetric worldline action of the massive $N=2$
superparticle  in $D=3$ \footnote{To avoid a possible confusion, recall that
$N=2$,  $D=3$ Poincar\'e supersymmetry has 4 fermionic generators
and it is just
$N=4$ superalgebra with two central charges from the $d=1$ perspective.}.

Note that this action was deduced earlier in \cite{gauntY} using
a different approach. It was demonstrated there that the corresponding
component action is related through a field redefinition to a
gauge-fixed form of the relevant Green-Schwarz type action (in the static
gauge  and with $\kappa $ symmetry fully used to gauge away two
out of four target spinor coordinates).

In the considered case, like in the previous one, there exists
one more possibility for the Goldstone superfield action,
with $\psi,\bar\psi$ being basic chiral superfields unrelated
to any additional bosonic ones.
In this case we once again end up with a $d=1$ Volkov-Akulov action
for the complex Goldstone fermionic field, such that it is reduced
to the free action via the suitable field redefinition.
\setcounter{equation}0
\section{N=4 $\rightarrow$ N=1  PBGS}
The conventional way of dealing with the PBGS phenomenon is to start from
some supersymmetry algebra, to construct its appropriate nonlinear
realization and then to look for the invariant action. As was already
mentioned in Introduction, the nonlinear realization formalism
is not too helpful in what concerns the last problem. Also,
there are no clear principles as to which version of the given
Poincar\'e supersymmetry should be actually chosen as the point
of departure: with or without central charges, how many such charges
should be taken into account, etc. For instance, an attempt to describe
$1/4$ breaking of $N=1$ Poincar\'e supersymmetry  in $d=11$ along
the standard lines \cite{bik} leads to very strong constraints on some of
the involved Goldstone superfields. Similar difficulties
come out when trying to apply the standard formalism for describing the
$1/4$ PBGS pattern $N=2, d=4 \rightarrow N=1, d=3$ \cite{ikunp}.
In this and next Sections we adhere to a somewhat different point of view
already accounted for in the previous Section. Namely, we start with some
``probe'' linear representation of the spontaneously broken supersymmetry
realized on a set of world-line superfields involving the needed
number of Goldstone fermions. Then we restore the full structure of
the supersymmetry algebra by studying the closure of supersymmetry
transformations. After this we express the linearly transforming
superfields in terms of the Goldstone ones, applying the procedure
explained in the previous Section (or its modifications). Finally,
we construct the  invariant off-shell Goldstone superfield action
like in the previous cases as a world-line superspace
integral of the appropriate  $d=1$ superfield component of the
original linear representation.

We firstly apply this general strategy to the case of
$N=4 \rightarrow N=1 $ PBGS. Our goal is to construct a
$N=1, d=1$ superfield
action which  would respect three extra spontaneously
broken supersymmetries.
Thus, the minimal multiplet should include at least three $N=1$ fermionic
Goldstone superfields $\psi^i(t,\theta) (i=1,2,3)$. Without
loss of generality these superfields can be chosen to form a triplet with
respect to some $SO(3)$ automorphism group. As their basic property, they
should have inhomogeneous transformation laws with respect to the broken
supersymmetries, i.e. their transformations should start with the
corresponding Grassmann parameters. One can
check that this requirement is met in a minimal way and the algebra
of transformations gets closed at cost of adding one additional
fermionic $N=1$ superfield $\Upsilon(t,\theta)$. The transformations
under the broken supersymmetries read
\be \label{n4tr1}
\delta \psi^i = \epsilon^i\left(1
-\cD\Upsilon\right)-  \varepsilon^{ijk}\epsilon^j \cD \psi^k \;, \qquad
\delta\Upsilon=\epsilon^i \cD \psi^i~. \ee
They form the following algebra
\be
\left\{ Q, Q \right\} = 2P\; , \qquad
\left\{ S^i, S^j \right\} = 2\delta^{ij}P\; , \qquad
\left\{ Q,S^i \right\} = 0\;. \label{susy4a}
\ee

The fermionic $N=1, d=1$ superfields contain no bosonic degrees
of freedom of physical dimension. Since we wish to have a
superparticle model, with the world-line scalar $N=1$ multiplets
containing such bosonic fields, we are led to introduce bosonic
superfields $v^i$
\be
\psi^i = \frac{1}{2}\cD v^i \;, \qquad
\left( {\bar\psi}{}^i=\psi^i\;, \; {\bar v}{}^i=-v^i \right) \;.
\ee
Their transformation properties can be extracted from
\p{n4tr1} :
\be\label{n1tr2}
\delta v^i = -2\epsilon^i\left(\theta -\Upsilon\right)+
 \varepsilon^{ijk}\epsilon^j \cD v^k \;, \qquad
\delta\Upsilon=\frac{1}{2}\epsilon^i \partial_t v^i \; .
\ee
Like in the $N=4 \rightarrow N=2$ case, due to  the explicit  presence of
$\theta$ in the transformations \p{n1tr2}, the anticommutators of
the manifest $Q$ and spontaneously broken $S^i$ supersymmetry generators
acquire an active central charges $Z^i$ in the right-hand side. Finally,
the basic anticommutation relations extracted from the above superfield
variations read
\be
\left\{ Q, Q \right\} = 2P\; , \qquad
\left\{ S^i, S^j \right\} = 2\delta^{ij}P\; , \qquad
\left\{ Q,S^i \right\} =2 Z^i~. \label{susy4}
\ee
The central charge generators act as pure shifts of $v^i$, suggesting
the interpretation of $v^i$ as Goldstone superfields
parametrizing transverse directions in a four-dimensional space
where $Z^i, P$ act as the translation operators.

Surprisingly, the superalgebra \p{susy4} cannot be interpreted as a
dimensional reduction of the standard $N=1$ Poincar\'e superalgebra
in $d=4$, with $Z^i, P$ being the components of full $4$- momentum.
Moreover, it also cannot be recovered from any one - or
two-central  charges extension of the $D=3$ or $D=2$
Poincar\'e superalgebras with the relevant Lorentz groups
($SO(1,2)\sim  SL(2,R)$ and $SO(1,1)$) as the automorphism ones.
Indeed, the only automorphism group of \p{susy4} is $SO(3)$ with
respect to which both  odd and even generators are split into a
singlet and triplet.

Nevertheless, it is still possible to interpret \p{susy4} in
the dimensional reduction language. For this one should proceed
not from the standard $N=1$, $D=4$ super Poincar\'e algebra, but from its
extension by tensorial  central charges \cite{1,ferrp,blstenz}.
The generators $Z^i$ turn out to partly come from these central charges
and partly from the extra components of  $4$-momentum. The precise
correspondence is given  in Appendix.

Let us turn to the issue of constructing invariant action for the
system under consideration.
Like in the previous case (Sect. 2), we can use $\Upsilon(v)$ as a Lagrangian
density,
\be\label{n4action1}
S_v= \int dt d\theta \Upsilon(v) \;,
\ee
in view of its transformation property \p{n4tr1}.
Then the main question is how to covariantly express $\Upsilon $ in
terms of $\psi^i$ and, further, $v^i$. One could use just the method of the 
previous
Section. However, it turns out that in the present case it is easier
to perform a direct construction of $\Upsilon $.

The idea of this construction is rather simple. In the case at
hand there is only one  non-nilpotent bosonic dimensionless object $
X=\cD\psi^i\cD\psi^i$. Therefore the general ansatz for the superfield
$\Upsilon$ will contain arbitrary functions of $X$ only.
Moreover, the unique objects having positive (one half)
dimension (in the length units) are the
spinor superfields $\psi^i$. This allows one to write the general
ansatz for $\Upsilon$ as
\bea
\Upsilon & = & \psi^i\cD\psi^i A + \psi^{2i}\psi^i_tB +\psi^{2i}\cD\psi^i
    \psi^j_t\cD\psi^j C+\psi^i\psi^i_t\psi^j\cD\psi^jE \nn
  && + \psi^3\cD\psi^i_t\cD\psi^i F +\psi^3\psi^{2i}_t\cD\psi^i G~, 
\label{n1action2}
\eea
where $A,B,\ldots,G$ are as yet undetermined functions of $X$,
and we use the following notations
\be
\psi_t^i = \partial_t \psi^i\;, \qquad
\psi^{2i} =\varepsilon^{ijk}\psi^j\psi^k \;, \qquad
\psi^3= \varepsilon^{ijk}\psi^i\psi^j\psi^k~.
\ee
Now, using \p{n4tr1}, \p{n1action2} we can write $\delta\psi^i$ in
terms of $\psi^i$ only. Then we can explicitly evaluate
$\delta\Upsilon$ and then require it to be equal to $\epsilon^i \cD
\psi^i$ in accordance with the transformation law \p{n4tr1}.
After rather lengthy calculations we  get the system of {\it algebraic}
equations for the functions $A,\ldots, G$ \bea\label{coeff}
&& \left( 1-XA\right)A=1 \Rightarrow
A=\frac{2}{1+\sqrt{1-4\cD\psi^i\cD\psi^i}}~, \nn
&&B=\frac{A^2}{2(A-2)},\,C=-\frac{A^4}{2(A-2)},\,E=\frac{A^3}{A-2},\,
F=\frac{A^3(A-4)}{6(A-2)^2},\, G=-\frac{A^5(A-4)}{6(A-2)^2}\,.
\eea
Thus the integral \p{n4action1} with $\Upsilon$ defined by
\p{n1action2}, \p{coeff} provides us with the action for the system
realizing the $N=4\rightarrow N=1$ PBGS pattern.

In fact we can greatly simplify this action. First, let us note
that the $B$ and $C$ terms in \p{n1action2} can be absorbed
(up to full $t$- and $\cD$- derivatives) into the $F$ term. As
the second step, all  the remaining $E$, $F$, $G$ terms can be
reduced to the single $A$ term, redefining the superfields
$v^i$ as follows
\be
v^i\rightarrow \phi^i = v^i+ \psi^3 \varepsilon^{ijk}\psi^{j}_t \cD\psi^k
H_1+\psi^3 \psi^i_t H_2+     \varepsilon^{ijk}\psi^{2j}\cD \psi^k H_3 \;,
\ee
where $H_1,\; H_2, \; H_3$ are some functions of $X$. These functions
can be given explicitly, but to know the precise expressions is in fact
needless for our purpose. The main point is that the action in terms
of the redefined bosonic superfield $\phi^i$ takes the very simple form
\be\label{action1f}
S_\phi=\int dt d\theta \frac{2\xi^i\cD\xi^i}{1+\sqrt{1-4\cD\xi^j\cD\xi^j}}\;,
\qquad
\xi^i \equiv \frac{1}{2}\cD\phi^i~.
\ee
Of course, the transformation properties of the new superfields
$\xi^i$, $\phi^i$ essentially differ from \p{n4tr1}, \p{n1tr2}, but
the action is guaranteed to be invariant by the above construction.

Thus we have found the correct Goldstone superfields action
describing the PBGS pattern $N=4\rightarrow N=1$.

Let us end this section by noting that the bosonic core of the action
\p{action1f}
\be
S_\phi^{bos}=\frac{1}{2}\int dt  \left( 1- \sqrt{1-\partial_t \phi^i
  \partial_t \phi^i}  \right)
\ee
is just the standard massive $D=4$ particle action in the static gauge.
It is known to exhibit the hidden nonlinearly realized $D=4$ Lorentz
symmetry $SO(1,3)$. One can ask why this symmetry is present in this
action whereas the supersymmetry algebra \p{susy4} from which
we have started respects no such an automorphism. The answer is that the
explicit  breaking of this Lorentz symmetry occurs just in the
fermionic terms of the component action. This becomes transparent
in the Green-Schwarz formulation of the same system. Now we turn
to describing such a formulation.
\setcounter{equation}0
\section{Target space action with one $\kappa$-supersymmetry}
To clarify the situation with $N=4 \rightarrow N=1$ PBGS, in this section
we construct the target space action which possesses only one
$\kappa$-supersymmetry and reduces to the action \p{action1f} in a
fixed gauge.

We shall deal with the $N=4$ superalgebra \p{susy4}. In accordance
with the standard strategy of constructing Green-Schwarz type
actions (see \cite{{aluk},{town},{town2},{town3},{gauntY}} for the
case of massive superparticles) let us
introduce bosonic $X^0(t),Y^i(t)$ and fermionic  $\Theta (t), \Psi^i (t)$
$d=1$ fields, the coordinates of a target $N=4$ superspace. They have the
standard transformation properties under $N=4$ supersymmetry  \p{susy4}
\be\label{GStr}
\delta X^0=-\epsilon \Theta-\epsilon^i\Psi^i\; ,\qquad
\delta Y^i=-\epsilon^i\Theta-\epsilon\Psi^i\;
,\qquad \delta\Theta=\epsilon\;,\qquad
\delta\Psi^i=\epsilon^i~.
\ee
Next we construct the supersymmetric invariants $\Pi^0,\Pi^i$
\be\label{GSPi}
\Pi^0={\partial_t X}{}^0+\Theta\partial_t\Theta+
  \Psi^i\partial_t\Psi{}^i~, \qquad
\Pi^i={\partial_t Y}{}^i-\partial_t \Theta \Psi^i+
 \Theta {\partial_t \Psi}{}^i~.
\ee

After some guess-work, the target sigma-model action invariant
under the global target space supersymmetry \p{GStr}, local
$t$ reparametrizations and one local fermionic $\kappa$
symmetry was found to have the following almost unique form
\be\label{GSaction}
S_{gs}=-\int dt \sqrt{\Pi^0\Pi^0 - \Pi^i\Pi^i} -\alpha  \int dt
     \left( \Theta\partial_t \Theta - \Psi^i\partial_t \Psi{}^i\right) \;,
\ee
where $\alpha = \pm 1$. The latter ambiguity in the sign amounts to
the fact that the first and second terms in the lagrangian density
in \p{GSaction} behave in  different way under the reflection $t \rightarrow
-t$: first one is invariant while the second changes its sign.
The cases with $\alpha = \pm 1$ are related by this reflection, so
without loss of generality one can choose, e.g., $\alpha = 1$.

The $\kappa$-symmetry transformations are given by
\bea
&& \delta\Theta = \kappa~, \qquad
\delta\Psi^i= \kappa\; \frac{\Pi^i}{\Pi^0 - \alpha \sqrt{\Pi^0\Pi^0
- \Pi^i\Pi^i} } \nn
&& \delta X^0=-\Theta\delta\Theta-\Psi^i\delta\Psi^i ,\qquad
\delta Y^i=-\Psi^i\delta\Theta-
 \Theta\delta\Psi^i~,\label{GSkappa}
\eea
where $\kappa(t)$ is an arbitrary fermionic gauge parameter. It is
straightforward to check the invariance of \p{GSaction} under these
transformations.

We claim that the action \p{GSaction}
possesses only one $\kappa$-supersymmetry and therefore provides
another, ``space-time''  realization of the same $N=4\rightarrow N=1$ PBGS
phenomenon. Let us prove that \p{GSaction} indeed possesses no any other
local fermionic symmetry apart from $\kappa$-symmetry \p{GSkappa}.

To this end, we need to study the algebra of the constraints in the
Hamiltonian formalism. We first introduce the einbein $e(t)$ and
rewrite the action \p{GSaction} (with $\alpha = 1$) as
\begin{equation} S_{gs}=\int dt
L=-\int dt\left[ \frac{1}{2e}\left( \Pi^0\Pi^0-\Pi^i\Pi^i\right)+
 \frac{e}{2}\right]
-\int dt\left( \Theta\partial_t \Theta-\Psi^i\partial_t \Psi^i\right)\;.
\end{equation}
Then one computes
canonically conjugated variables
\begin{eqnarray}
&&P^0 =\frac{\partial L}{\partial\dot  X^0}=-\frac{\Pi^0}{e}\;,\quad
P^i =\frac{\partial L}{\partial\dot  Y^i}=\frac{\Pi^i}{e}\;, \quad
P_e=\frac{\partial L}{\partial\dot  e}=0\;,\nonumber\\
&&\Omega =\frac{\partial L}{\partial\dot  \Theta}=
  \left( \frac{\Pi^0}{e}+1\right)\Theta-
\frac{\Pi^i}{e}\Psi^i\;,\quad
\Omega^i=\frac{\partial L}{\partial\dot  \Psi^i}=
 \left( \frac{\Pi^0}{e}-1\right)\Psi^i-
 \frac{\Pi^i}{e}\Theta \;.
\end{eqnarray}
The canonical hamiltonian reads
\begin{equation}
H=P^0\partial_t  X^0+P^i\partial_t  Y^i+\partial_t \Theta\Omega+\partial_t 
\Psi^i\Omega^i-L
=-{e\over 2}(P^0P^0-P^iP^i-1)~.
\end{equation}
There is one primary bosonic constraint, $P_e$, and four fermionic constraints
which we denote by
\begin{equation}
\tau^0=\Omega+ (P^0-1)\Theta+P^i\Psi^i~, \qquad
\tau^i=\Omega^i+(P^0+1)\Psi^i+P^i\Theta\;.
\end{equation}
When taking the Poisson bracket of the primary bosonic constraint with the 
canonical
hamiltonian, we obtain the secondary bosonic constraint
\begin{equation}  P^0P^0-P^iP^i=1  \end{equation}
We now have to determine which of the fermionic
constraints $\tau^\mu = (\tau^0, \tau^i)$ are first class,
and thus generate  gauge symmetries, and which are second class. We compute
the matrix of the Poisson brackets of the fermionic constraints
\begin{equation}
\{\tau^\mu,\tau^\nu\}=C^{\mu\nu}~, \qquad
C=2\left(\begin{array}{cc}P^0-1 & \vec P^t \\
\vec P & (P^0+1)\mbox{\bf 1}
\end{array}\right)~,
\end{equation}
where {\bf 1} is the three dimensional unit matrix. The eigenvalues of the 
matrix $C$ are easily
computed to be $P^0+1$, $P^0+1$, $P^0+\sqrt{\vec P^2+1}$,
$P^0-\sqrt{\vec P^2+1}$, On the constraint surface, the last of these 
eigenvalues vanishes,
and the other three remain non zero. Thus, there is one and only one first
class constraint which may be chosen to be
\begin{equation}
\kappa=\tau^0-{1\over P^0+1}\vec P\vec\tau.
\end{equation}
Its Poisson brackets with the constraints read
\begin{equation}
\{\kappa,\tau^0\}={2(P^0P^0-P^iP^i-1)\over P^0+1}~,
\qquad \{\kappa,\tau^i\}=0~.
\end{equation}
The constraint $\kappa$ generates the unique local fermionic symmetry
\p{GSkappa} through the Poisson bracket.

It is natural to expect, by analogy with the $1/2$ PBGS examples
\cite{{town},{town2}}, that the static gauge $( X^0=t, \Theta=0)$ form of
\p{GSaction} coincides, modulo field redefinitions, with the component
action following from the world-line superfield one \p{action1f}.
Now we shall show that this is indeed the case.

In the static gauge the action \p{GSaction} (with $\alpha = 1$) reads
\be\label{GSaction3}
S_{gs}=-\int dt \left[ \sqrt{ \left(1+ \Psi^i{\partial_t  \Psi}{}^i\right)^2-
   {\partial_t  Y}{}^i{\partial_t  Y}{}^i }- \Psi^i{\partial_t
\Psi}{}^i\right]~.
\ee
One can reduce it to the form
\be\label{lastGS}
S_{gs}=-\int dt \left[ \sqrt{ 1- {\partial_t   Y}{}^i {\partial_t  Y}{}^i} -
  {\widetilde\Psi}{}^i{\partial_t {\widetilde\Psi}}{}^i \right]~,
\ee
where
\bea
{\widetilde \Psi}{}^i &=& a_1\Psi^i \left[ 1 +a_2 ( \Psi{\partial_t  \Psi})+
  a_3 (\Psi{\partial_t  \Psi})^2 \right]~,
\quad
a_1^2=1+\frac{1}{\sqrt{ 1-{\partial_t   Y}{}^i {\partial_t  Y}{}^i}}~, \nn
a_2&=&\frac{{\partial_t   Y}{}^i {\partial_t  Y}{}^i}{4a_1^2
\left( 1-{\partial_t   Y}{}^j {\partial_t
  Y}{}^j\right)^{3/2} }~, \quad
a_3=
\frac{{\partial_t   Y}{}^i
{\partial_t  Y}{}^i}{4a_1^2\left( 1-{\partial_t   Y}{}^j {\partial_t
  Y}{}^j\right)^{5/2} } -\frac{a_2^2}{2}~.
\eea
{}On the other hand, the action \p{action1f} after integration over
$\theta$ and properly rescaling fermions can be rewritten as follows:
\be\label{nonGS1}
S_\phi=\int dt\left[ \frac{\frac{1}{2} {\partial_t  \phi}{}^i{\partial_t  
\phi}{}^i }{ 1+\sqrt{1 - {\partial_t
\phi}{}^j{\partial_t  \phi}{}^j} } -\xi^i{\partial_t  \xi}{}^i 
-\frac{\xi^i{\partial_t \phi}{}^i
 {\partial_t \xi}{}^j{\partial_t  \phi}{}^j }
{ (1+\sqrt{1 - {\partial_t  \phi}{}^j{\partial_t  \phi}{}^j} )\sqrt{1 - 
{\partial_t
\phi}{}^j{\partial_t  \phi}{}^j}} \right]~.
\ee
By passing to the new variables $\phi^i, {\widetilde\xi}{}^i$
\be\label{nonGS2}
\xi^i={\widetilde\xi}{}^i\left( 1+
A{\partial_t \phi}{}^j{\widetilde\xi}{}^j\partial_t
  (\partial_t \phi^k{\partial_t {\widetilde\xi}}{}^k) \right)
\ee
the action \p{nonGS1} can be put in the form
\be\label{lastnonGS}
S_\phi=\int dt \left[\frac{1}{2}- \frac{1}{2}\sqrt{ 1- {\partial_t {\widetilde 
\phi}}{}^i
{\partial_t {\widetilde \phi}}{}^i      } +
\psi^i{\partial_t {\psi}}{}^i  \right]~.
\ee
Here
\be
{\widetilde\phi}{}^i=\phi^i+b_1{\widetilde\xi}{}^i
 {\partial_t {\widetilde\phi}}{}^j{\widetilde\xi}{}^j~, \qquad
\psi^i= {\widetilde\xi}{}^i+ b_2
  {\partial_t \phi}^i{\partial_t \phi}{}^j{\widetilde\xi}{}^j~,
\ee
and $A,b_1,b_2$ are defined by simple algebraic equations
\bea
&&b_2^2({\partial_t \phi})^2=
\frac{2b_1}{\sqrt{ 1-({\partial_t  \phi})^2}(1- \sqrt{ 1-({\partial_t
\phi})^2}) }~, \;A=-\frac{b_1^2}{\sqrt{ 1-({\partial_t  \phi})^2}(1- \sqrt{
1-({\partial_t  \phi})^2})}~, \nn
&& 2b_2+b_2^2({\partial_t \phi})^2=
\frac{4}{\sqrt{ 1-({\partial_t  \phi})^2}(1- \sqrt{ 1-({\partial_t
\phi})^2})}~.
\eea
It can be easily seen  that after rescaling of
${\widetilde\psi}{}^i$ by  $1/\sqrt{2}$ the
action \p{lastnonGS} reduces to \p{lastGS} up to an overall constant.

Finally, we note that in terms of the invariants \p{GSPi} the Green-Schwarz
action \p{GSaction} looks as if it possess $D=4$ Lorentz invariance.
Indeed, in the limit of vanishing fermions $X^0$, $Y^i$ and, hence,
$\Pi^0, \Pi^i$ can be combined into a $D=4$ vector. However, the
fermionic terms in \p{GSPi} break this ``would-be'' Lorentz symmetry
down to $SO(3)$.
\setcounter{equation}0

\section{N=8 $\rightarrow$ N=2 PBGS}
In this section we will consider  two examples of
$N=8 \rightarrow N=2$ PBGS.

In the previous sections we described the procedure which helps
to define the action for the given PBGS pattern if the proper
realization of broken supersymmetries is known. To construct a
superparticle model which would exhibit $N=8 \rightarrow N=2$ PBGS
we should, before all, examine how 6 broken
supersymmetries could be realized on a set of $N=2, \;d=1$ superfields.
We succeeded in finding out two such realizations.

\subsection{Case I}

The first realization is a more or less straightforward generalization of
the $N=4\rightarrow N=1$ case. The basic set of $N=2,d=1$ superfields
includes seven bosonic superfields: a general real superfield $\Phi$ and
two conjugated triplets of chiral-anti-chiral superfields
${\bar v}{}^i, v^i$:
$$D v^i = \Db {\bar v}{}^i =0\;, \quad i=1,2,3 \;.$$
The broken supersymmetry transformations of these superfields read
\bea
&& \delta v^i = -2\left( \tb - D \Phi\right) \epsilon^i
+\varepsilon^{ijk}{\bar\epsilon}{}^j D {\bar v}{}^k~, \quad
 \delta {\bar v}{}^i = 2\left( \theta +  \Db \Phi\right) {\bar\epsilon}{}^i
+\varepsilon^{ijk}\epsilon^j \Db v^k , \nn
&& \delta \Phi = \frac{1}{2}\left( \epsilon^i D {\bar v}{}^i + 
{\bar\epsilon}{}^i\Db v^i \right) .
\label{n83}
\eea
Together with the manifest
supersymmetry, they form the algebra  with six central charges $Z^i,{\bar
Z}{}^i$
\be
 \left\{ Q,{\bar Q} \right\} = P\;,\quad
\left\{ S^i,{\bar S}{}^j \right\} = \delta^{ij} P\;, \quad
 \left\{ Q, S^i \right\} =2 Z^i\;, \quad
\left\{ {\bar Q}, {\bar S}{}^i \right\} = 2{\bar Z}{}^i\;. \label{n84}
\ee
The fermionic chiral superfields defined by
$$\psi^i = -\frac{1}{2} \Db v^i\; , \qquad  \bar\psi{}^i = \frac{1}{2} D {\bar 
v}{}^i $$
are transformed under \p{n83} as
\bea
\delta \psi^i = \left( 1 - \Db D\Phi\right) \epsilon^i
+\varepsilon^{ijk}{\bar\epsilon}{}^j \Db {\bar \psi}{}^k ,\;
 \delta {\bar\psi}{}^i = \left( 1 + D \Db  \Phi\right) {\bar\epsilon}{}^i
+\varepsilon^{ijk}\epsilon^j D \psi^k , \;
\delta \Phi =\epsilon^i{\bar\psi}{}^i- {\bar\epsilon}{}^i\psi^i  . &&
\label{n81}
\eea
So they are Goldstone superfields corresponding to
the linear realization of six spontaneously broken supersymmetries
with the parameters $\epsilon^i, {\bar\epsilon}{}^i$. The bosonic
superfields ${\bar v}{}^i, v^i$ are the Goldstone ones associated
with the spontaneously broken central charges transformations.

An interesting feature of this realization
is the strange charges of the Goldstone superfields with respect
to the $U(1)$ automorphism group of the manifest supersymmetry algebra.
Indeed, from \p{n81} one finds the relation between the charges
of $\psi^i$  and $\theta$
\be
q_{\psi} = {1\over 3}q_{\theta}\;.
\ee
Correspondingly, for the charge of  $v^i$ we have
$$
q_{v} = -{2\over 3}q_{\theta}\;.
$$
One should ascribe similar fractional charges, of course, to the
spontaneously broken supersymmetry and central charge generators
in \p{n84}.

Once again, the superfield $\Phi$, in accord with its transformation
properties, can be chosen as the Lagrangian density describing
this  PBGS pattern.
The straightforward application of the method \cite{{ikap1},{ikap2}} for
expressing $\Phi$ in terms of the Goldstone superfields
$\psi^i,{\bar\psi}{}^i$ yields a rather complicated system of equations.
It can be rather easily solved in the limit of vanishing fermions,
yielding the static gauge action for a massive particle in a
seven-dimensional space-time as the bosonic part of the full
superfield action
\be\label{n85}
S_v^{bos}=\frac{1}{2}\int dt \left( 1- \sqrt{1+ \partial_t v^i \partial_t{\bar 
v}{}^i}
 \right)~.
\ee
We have found the full action in terms of $N=2$ superfields as well.
It does not look very illuminating, so we do not give it here.

Surprisingly, the target space GS formulation for this case
is very similar to the case of $N=4\rightarrow N=1$ PBGS.
The full action for the physical worldline multiplet can be immediately
extracted from this formulation.

As a first step, we define the standard realization of N=8
superalgebra \p{n84} in the superspace with seven bosonic
$X^0,Y^i,{\bar
Y}{}^i$ and eight fermionic $\Theta,{\bar\Theta},\Psi^i,{\bar\Psi}{}^i$
coordinates: \bea\label{n86}
&& \delta \Theta = \epsilon~, \quad \delta \Psi^i = \epsilon^i~, \quad
       \delta Y^i=-2\epsilon^i \Theta~, \quad
 \delta \bar\Theta = \bar\epsilon~, \quad
\delta \bar\Psi{}^i = \bar\epsilon{}^i~, \quad
       \delta {\bar Y}{}^i=2\bar\epsilon{}^i \bar\Theta \;  , \nn
&& \delta X^0= -\frac{1}{2}\left( \epsilon\bar\Theta+\bar\epsilon\Theta+
    \epsilon^i{\bar\Psi}{}^i+{\bar\epsilon}{}^i \Psi^i \right)~.
\eea
Using the supersymmetric invariants
\bea\label{n87}
&&\Pi^0={\partial_t  X}{}^0+\frac{1}{2}\left( \Theta {\partial_t {\bar\Theta}}
+{\bar\Theta}{\partial_t \Theta} +\Psi^i{\partial_t {\bar \Psi}}{}^i+
{\bar\Psi}{}^i{\partial_t \Psi}{}^i\right)\;,\nn
&&\Pi^i= {\partial_t  Y}{}^i+2{\Psi^i}{\partial_t \Theta}~, \quad
{\bar\Pi}{}^i \equiv \bar{\left( \Pi^i\right)} =
-{\partial_t {\bar{Y}}}{}^i+2{{\bar\Psi}{}^i}{\partial_t {\bar\Theta}}~.
\eea
we can construct the unique action
\be\label{n88}
S_{gs}=-\int dt \sqrt{ \Pi^0\Pi^0 - \Pi^i{\bar{\Pi}}{}^i } +
   \int dt \left( \Theta{\bar{\partial_t {\Theta}}} -\Psi^i {\partial_t 
{\bar\Psi}}{}^i
   \right)~,
\ee
with two $\kappa$-supersymmetries:
\bea\label{n89}
&& \delta X^0= -\frac{1}{2}\left( \bar\Theta\delta\Theta+
  \Theta\delta{\bar\Theta} +
    {\bar\Psi}{}^i\delta\Psi^i+\Psi^i\delta{\bar\Psi}{}^i \right)~,
\;  \delta Y^i = -2\Psi^i\delta\Theta~, \;
\delta {\bar Y}{}^i = 2{\bar\Psi}{}^i\delta{\bar\Theta}\;, \nn
&& \delta \Psi^i = \frac{ \Pi^i\delta{\bar\Theta} }
      {  \Pi^0+\sqrt{  \Pi^0\Pi^0 - \Pi^i{\bar{\Pi}}{}^i } }~, \;
\delta {\bar\Psi}{}^i = \frac{ {\bar\Pi}{}^i\delta{\Theta} }
      {\Pi^0+\sqrt{  \Pi^0\Pi^0 - \Pi^i{\bar{\Pi}}{}^i } }\; .
\eea
The Hamiltonian analysis, which repeats the basic steps of the
analysis in the $N=4\rightarrow N=1$ case, shows that there are
no further gauge fermionic symmetries in the action \p{n88}.

In the static gauge, $ X^0=t, \Theta=0$, the action \p{n88} takes the very
simple form
\be\label{n810}
S_{gs}=-\int dt \left[ \sqrt{ \left(1+\frac{1}{2}\Psi^i{\partial_t
{\bar\Psi}}{}^i+
\frac{1}{2} {\partial_t \Psi}{}^i{\bar\Psi}{}^i \right)^2 +
  {\partial_t  Y}{}^i{\partial_t {\bar{Y}}}{}^i } +\Psi^i{\partial_t 
{\bar\Psi}}{}^i \right]
\;.
\ee
We are still not aware of the equivalency transformation from
this action to the PBGS action. We expect such a transformation
to exist like in the other known PBGS cases, though its precise form
can be rather complicated in view of complexity of the full PBGS action.

\subsection{Case II}

The second realization of $N=8, d=1$ supersymmetry with
six spontaneously broken supersymmetries can be constructed in terms
of general bosonic $N=2$ superfield $\Phi$ and six chiral and anti-chiral
Goldstone fermions  $\left\{ \psi_{\alpha},{\bar \psi}_\alpha, \xi,
\bar\xi\right\}, \alpha=1,2$ \be\label{n811}
\Db \psi_\alpha=\Db \xi=0~, \qquad D {\bar\psi}_\alpha= D {\bar\xi}=0\;,\;
\ee
which form two doublets and two singlets with respect to
$SO(2)$ automorphism group. The appropriate closed set of
the broken supersymmetry transformations reads
\bea\label{n812}
&&\delta\xi =\left(1+\Db D\Phi\right) \nu +
 \varepsilon_{\alpha\beta}{\bar\mu}_\alpha \Db{\bar\psi}_\beta~, \quad
\delta{\bar\xi} =\left(1-D\Db \Phi\right) {\bar\nu} +
  \varepsilon_{\alpha\beta} \mu_\alpha D \psi_\beta \;, \nn
&&\delta\psi_\alpha=\varepsilon_{\alpha\beta}\left( \bar\nu
   \Db{\bar\psi}_\beta +{\bar\mu}_\beta \Db{\bar\xi}\right)+
 \left(1-\Db D\Phi\right)\mu_\alpha \; , \nn
&&\delta{\bar\psi}_\alpha=\varepsilon_{\alpha\beta}\left(
 \nu D \psi_\beta+\mu_\beta  D \xi \right)+
   \left(1+ D\Db \Phi\right){\bar\mu}_\alpha \; , \nn
&&\delta\Phi= {\bar\nu} \xi-\nu{\bar\xi}-
   {\bar\mu}_\alpha\psi_\alpha+\mu_\alpha{\bar\psi}_\alpha\; .
\eea

To reveal the underlying central-charges extended supersymmetry
algebra and to gain physical bosonic fields, we need to pass as before
to  bosonic superfields. The minimal realization amounts
to expressing $\psi_\alpha$
through two real scalar superfields $u_\alpha$:
\be\label{n814}
\psi_\alpha=-\frac{1}{2}\Db u_\alpha~, \qquad {\bar\psi}_\alpha=\frac{1}{2} D 
u_\alpha\;.
\ee
To learn what kind of ``prepotential'' one should introduce
for the remaining Goldstone superfield $\xi$, let us examine
the relation between $U(1)$ charges of spinor superfields which
follows from \p{n812}
\be\label{n813}
q_{\xi}=-2q_{\psi}-q_D\;.
\ee
Here $q_D$ is the $U(1)$ charge of the covariant derivative $D$
($q_D = -1$ if one ascribes the charge $+1$ to $\theta$).
{}From this relation and \p{n814} it follows that
the $U(1)$ charges of $\psi$ and $\xi$  are equal to $-q_D$ and
$q_D$, respectively.
Recall, however, that $\psi$ and $\xi$ have same  chiralities (see
\p{n811}). Bearing this in mind, the only  way to introduce the bosonic
superfield $v$ for $\xi$ is to choose it complex and having
the $U(1)$ charge equal  to $-2q_D$
\be\label{n815}
\xi=-\frac{1}{2}\Db v~, \qquad {\bar\xi} = \frac{1}{2}D {\bar v}~,
\qquad Dv = \Db{\bar v} =0 \;.
\ee

In terms of the newly introduced bosonic superfields the
supersymmetry transformations take the form:
\bea\label{n816}
&&\delta v=-2\left(\tb+ D\Phi\right) \nu +
\varepsilon_{\alpha\beta}{\bar\mu}_\alpha D u_\beta~, \quad
\delta{\bar v} =2\left( \theta- \Db \Phi\right) {\bar\nu} +
  \varepsilon_{\alpha\beta} \mu_\alpha \Db u_\beta \;, \nn
&&\delta u_\alpha= \varepsilon_{\alpha\beta}\left(
 {\bar\nu} D u_\beta+\nu \Db u_\beta +
 {\bar\mu}_\beta D {\bar v} +\mu_\beta \Db v\right)+
   2\left(\theta+ \Db\Phi\right){\bar\mu}_\alpha -
   2\left({\bar\theta}- D\Phi\right)\mu_\alpha\; , \nn
&&\delta\Phi= -\frac{1}{2}\left( \nu D{\bar v}+{\bar\nu}\Db v-
   {\bar\mu}_\alpha \Db u_\alpha- \mu_\alpha D u_\alpha\right)\; .
\eea
Denoting the generators of the broken supersymmetry by
$S_\alpha, {\bar S}_\alpha$ and $S, {\bar S}$, and the generators of
the manifest $N=2$ supersymmetry by $Q,{\bar Q}$, one can write the
full supersymmetry algebra pertinent to this case as
\bea\label{n817}
&& \left\{ Q,{\bar Q} \right\}= \left\{ S,{\bar S} \right\}=P~,\;
\left\{ S_\alpha,{\bar S}_\beta \right\}=\delta_{\alpha,\beta}P~, \;
\left\{ Q,{\bar S} \right\}=2{\bar Z}~, \;
 \left\{ {\bar Q}, S \right\}=2Z~,\nn
&& \left\{ Q, {\bar S}_\alpha \right\}=2Z_\alpha~, \;
 \left\{ {\bar Q}, S_\alpha \right\}=2Z_\alpha~, \;
\left\{ S, {\bar S}_\alpha \right\}=
 2\varepsilon_{\alpha\beta}Z_\beta~,\;
 \left\{ {\bar S},S_\alpha \right\}=
 2\varepsilon_{\alpha\beta}Z_\beta~.
\eea

Once again, we can take the superfield $\Phi$ as the Lagrangian
density and the main problem is to covariantly express $\Phi$ in terms of
the Goldstone fermions or Goldstone bosons. The method we
used in the previous cases works here as well, but it gives
rather complicated equations which we for the time being were unable to
solve explicitly. Nevertheless, we can try to find the bosonic part of the
action. To this end, we should keep in the superfields with
tilde (which are obtained by the finite broken supersymmetry
transformations of our basic superfields) only a few terms, namely,
the terms linear   in the fermionic superfields
and quadratic terms in $\tilde\Phi$: \be\label{n818}
\left( \begin{array}{c} \tilde\xi \\ {\tilde{\bar\xi}} \\
 {\tilde\psi}_\alpha\\
 {\tilde{\bar\psi}}_\alpha \end{array} \right)=
\left( \begin{array}{c} \xi \\ {\bar\xi} \\ \psi_\alpha\\
 {\bar\psi}_\alpha \end{array} \right)+
\left( \begin{array}{cccc}
1+X & 0 & 0 & \varepsilon_{\beta\gamma}\Db {\bar\psi}_\gamma   \\
0 & 1+X &  \varepsilon_{\beta\gamma}D \psi_\gamma & 0 \\
0& \varepsilon_{\alpha\gamma}\Db{\bar\psi}_\gamma &  (1-X)\delta_{\alpha\beta} &
  \varepsilon_{\alpha\beta}\Db{\bar\xi} \\
 \varepsilon_{\alpha\gamma} D \psi_\gamma &0 &
\varepsilon_{\alpha\beta} D \xi & (1-X)\delta_{\alpha\beta}
\end{array}\right)
\left( \begin{array}{c} \nu \\ {\bar\nu} \\ \mu_\beta \\
 {\bar\mu}_\beta  \end{array} \right) = 0~,
\ee
\be\label{n819}
\tilde\Phi=\Phi+\frac{1}{2}\left(
{\bar\nu}\xi-\nu{\bar\xi}+
\mu_\alpha{\bar\psi}_\alpha\ - {\bar\mu}_\alpha\psi_\alpha \right) = 0 \;,
\ee
where $X=\Db D\Phi$.

Now, substituting $( \nu , {\bar\nu} , \mu_\alpha , {\bar\mu}_\alpha )$
from \p{n818} in \p{n819}, hitting
both sides of \p{n819} by $\Db D$ and omitting the terms with
fermions we get the following equation for the bosonic part of the
action which we denote by $X$:
\be\label{n820}
(X^2-X+a)(X^2+a-1)+2D\xi\Db{\bar\xi} = 0 \; ,
\ee
or, equivalently,
\be\label{n821}
(X^2-2X+1+a)(X^2+X+a) -2  \Db \psi_\alpha D{\bar\psi}_\alpha  = 0 \; .
\ee
Here, $$a= D\xi\Db{\bar\xi}+  \Db \psi_\alpha D{\bar\psi}_\alpha  \;.$$
The general solution of these equations exists (we are interested
in that one which goes to zero when all fields are put equal to zero),
but it looks too complicated to explicitly present it here.
In the two limits,  $\psi_\alpha=0$ or $\xi=0$, it takes the familiar
form of the static gauge actions of massive particles moving on
some three-dimensional target manifolds
\be\label{n822}
S_v^{bos}=\frac{1}{2}\int dt \left( 1-\sqrt{1+ {\partial_t  v}{\partial_t {\bar 
v} }}
        \right)~, \qquad
S_u^{bos}=\frac{1}{2}\int dt \left( 1-\sqrt{ 1+
       {\partial_t  u}_\alpha {\partial_t  u}_\alpha}
       \right)\;.
\ee
In the generic case there is a non-trivial cross-interaction between
the bosonic fields appearing in \p{n822}. It can hopefully be interpreted
in the geometric language of intersection of the trajectories of two
different superparticles, with the physical worldline
scalar multiplets represented by the superfields $u_\alpha$ and
$v, \bar v $, respectively.

The fact that the bosonic part of the action cannot be written in
the standard form strongly obscures the construction of the GS
formulation for this case. Up to now we have not succeeded to find a
manageable form even for the bosonic part of the hypothetical
target superspace GS action in the case at hand (though the WZ term
seems to be a direct generalization of those for the previously
studied cases). We believe that the better understanding of this case
would be  helpful for studying  the $1/4$ PBGS systems with
higher-dimensional worldvolumes. Such systems hopefully correspond
to the new types of superbranes.

\section{Conclusions}
In this paper we presented, for the first time, the manifestly worldline
supersymmetric superparticle actions exhibiting hidden spontaneously
broken supersymmetries the number of which is four times the number of
the linearly realized manifest ones. We treated in detail the case
of $N=4 \rightarrow N=1$ partial breaking and discussed some basic features
of the more complicated $N=8 \rightarrow N=2$ case. We proposed a
general systematic method of constructing actions for these and other PBGS
systems. For the first
case  and for one of the two versions of the second case we found the
appropriate GS-type  target superspace actions with one and two
$\kappa$-symmetries. The common unusual feature of the
superparticle systems considered is that their space-time
interpretation is possible only within the superspaces corresponding
to higher-dimensional supersymmetries with tensorial central charge
generators. The target space coordinates associated with some
combinations of these generators and extra components of the
translation operator
parametrize the transverse spatial bosonic
directions in these models. It would be of interest to understand
whether this is the general property of systems with fractional
PBGS. Another interesting problem is to see whether the $1/4$ PBGS
patterns studied here can be related to the existence of the
appropriate BPS solutions of higher-dimensional supergravities
along the line of refs. \cite{{ggpt},{at},{gh}}.

\section*{Acknowledgments}
We would like to thank A.~Kapustnikov, J.~Lukierski, D.~L\"{u}st, A.~Pashnev,
P.~Pasti,
C.~Preitschopf, D.~Sorokin,  M.~Tonin and B.~Zupnik for many useful
discussions. E.I. and S.K. thank
Laboratoire de Physique de l'ENS Lyon for the
hospitality during
the course of this work. This work was supported in part
by the PICS
Project No. 593, RFBR-CNRS Grant No. 98-02-22034, RFBR Grant No. 99-02-18417,
Nato Grant No. PST.CLG 974874 and INTAS Grants INTAS-96-0538, INTAS-96-0308.

\setcounter{equation}0
\def\theequation{A.\arabic{equation}}
\section*{Appendix: N=4, d=1 from N=1, D=4}
Here we demonstrate how $N=4, \; d=1$ supersymmetry algebra \p{susy4}
can be recovered from the $N=1$, $D=4$ algebra extended by tensorial
central charges.

This extended  superalgebra is defined by the following
anticommutation relations \cite{{ferrp},{blstenz},{gh}}
\bea
&& \{Q_\alpha, \; \bar Q_{\dot \beta} \} =
2\,(\sigma^\mu)_{\alpha\dot \beta}P_{\mu} = 2\,\delta_{\alpha\dot \beta}P_0
+ 2\,(\sigma^m)_{\alpha\dot \beta}P_m~, \label{A1} \\
&& \{Q_\alpha, \; Q_{\beta} \} = 2\,T_{(\alpha\beta)}\;, \qquad
\{\bar Q_{\dot \alpha}, \; \bar Q_{\dot \beta} \} =
2\,\overline{T}_{(\dot \alpha\dot \beta)} \label{A2}
\eea
Let us deviate from this standard manifestly $SL(2,C)$ covariant
notation by keeping manifest only $SU(2) \in SL(2,C)$. In such a notation
the dotted indices are simply upper case doublet $SU(2)$ indices.
E.g., the relation \p{A1} is rewritten as
$$
\{Q_\alpha, \; \bar Q^{\beta} \} = 2\,\delta_{\alpha}^\beta P_0  +
2\,(\sigma^m)_{\alpha}^\beta P_m \;. \label{A3}
$$
Next, we combine $Q$ and $\bar Q$ into doublets of some extra $SU(2)$
and pass to the quartet notation
\bea
Q_{\alpha i} = (Q_\alpha, - \bar Q_{\beta}), \qquad
\overline{Q_{\alpha i}} = \epsilon^{\alpha\beta}\epsilon^{ik}
Q_{\beta k}\;, \qquad (\epsilon_{12} = -\epsilon^{12} = 1)\;. \label{A4}
\eea
In terms of these generators the original algebra \p{A1}
is rewritten as
\bea
\{Q_{\alpha i},\; Q_{\beta k} \} = 2\,F_{(\alpha \beta)(ik)} +
2\,\epsilon_{\alpha\beta} \epsilon_{ik}P_0\;, \label{A5}
\eea
where
\be
F_{(\alpha \beta)(11)} = T_{(\alpha \beta)}~, \quad
F_{(\alpha \beta)(22)} = \overline{T}_{(\alpha\beta)}~, \quad
F_{(\alpha \beta)(12)} =
-(\sigma^m)_{(\alpha\beta)}P_m~. \label{A6} \ee
Note that this form of the algebra \p{A1}, \p{A2} reveals that the full
automorphism group of the latter is the 16-parametric
non-compact general linear group $GL(4,R)$. It contains, in parallel with
the $SO(4) \sim SU(2) \times SU(2)$ subgroup which is manifest
in the notation \p{A5}, also the original Lorentz group, of course.
The generators belonging to the coset of the latter over
the left manifest $SU(2)$ are constructed as direct products of
the generators of this  $SU(2)$ by $i(\sigma^3)_i^k$ and so act also
on the index $i$ of  $Q_{\alpha i}$.

In order to reproduce the algebra \p{susy4}, we need two more steps.

First, we should pass to the notation in which only
the diagonal $SU(2)$ in the product $SU(2) \times SU(2)$ realized
on $Q_{\alpha i}$ is manifest. Then we split $Q_{\alpha i}$ into a singlet and
triplet with respect to this diagonal $SU(2)$
\be
Q_{\alpha i} = \epsilon_{\alpha i} Q + i (\sigma^n)_{(\alpha i)}S^n~,
\qquad Q = \bar Q\;, \; S^n = \overline{(S^n)}~. \label{A7}
\ee
Similarly, the tensorial generator $F_{(\alpha \beta)(ik)}$ in \p{A5}
is split into a real totally symmetric 4-index tensor
$F_{(\alpha \beta ik)}$ (5 independent components),
a pseudo-real triplet
\be
F_{(\alpha i)} = {1\over 2}\;\{\epsilon^{\beta k}F_{(\alpha \beta)(ik)}
+ (\alpha \leftrightarrow i)\} \equiv -2i (\sigma^m)_{(\alpha i)} Z^m~,
\qquad  Z^m = \overline{Z^m}\;,
\label{A8}
\ee
and a real singlet
\be
F = {1\over 2}\epsilon^{\alpha i}\epsilon^{\beta k}F_{(\alpha \beta)(ik)}~.
\label{A9}
\ee
The algebra \p{A5} in this new basis can be rewritten as follows
\bea
 \{Q, \;Q\} = P_0
+  F~, \quad
 \{Q, \; S^n\} = 2\,Z^n~, \quad
 \{S^m, \; S^n\} = 2\,Z^{(mn)} + \delta^{mn}\,(P_0 - {1\over 3}F)~,
\label{A12}
\eea
where
$$
Z^{(mn)} = -{1\over 4}(\sigma^m)^{(\alpha i)}(\sigma^n)^{(\beta k)}
F_{(\alpha\beta ik)}~,  \qquad \mbox{Tr} \,Z  = 0 \;.
$$

We observe that \p{A12} coincide with \p{susy4} (up to the
rescaling $P= 1/2 \,P_0$) provided  that
\be
Z^{(mn)} = 0~, \qquad F=0\;. \label{A13}
\ee
These constraints are covariant under the manifest diagonal $SU(2) \sim
SO(3)$ which is thus identified with the $SO(3)$ automorphisms of
\p{susy4}. Eqs. \p{A13} can be shown to fully break the rest of
the automorphism group $GL(4,R)$ of the algebra \p{A1}, \p{A5},
which explains the absence of any extra automorphisms in \p{susy4}.

Eqs. \p{A13} amount to the following constraints on the original
bosonic generators
\be
T_{12} = P_1 - iP_2 \;, \quad  \overline{T}_{12} = -(P_1 + iP_2) \;,
\quad \overline{T}_{11} + T_{22} = 0 \;, \quad
T_{11} = \overline{T}_{22} = 0 \;, \quad P_3 = 0 \;,
\ee
leaving us just with four independent generators (together with $P$).
Taking account of these relations, the central charges $Z^i$ can
be identified with the following combinations
\be
Z^1 = P_2~, \qquad Z^2 = -P_1~, \qquad Z^3 = {i\over 4}
(\overline{T}_{11} - T_{22})~.
\ee
Note that the superalgebra of charges obtained via N\"other's
prescription from the PBGS action \p{action1f} differs from \p{susy4} by the
presence of non-zero constant generator $F$. It ensures
the relative shift of the translation generators in the $QQ$ and
$SS$ anticommutators in accord with the general reasoning of
ref. \cite{{hp},{hlp}}.

For the underlying algebras of other $1/4$ PBGS examples considered in
this paper one can also obtain a similar identification proceeding from
the most general extension of $N=2, \;D=4$ superalgebra by the appropriate
tensorial central charges (such extensions for the generic $N$
are presented, e.g., in \cite{{ferrp},{gh}}).

\end{document}